\documentclass[aps,prl,twocolumn,superscriptaddress,showpacs,floatfix,preprintnumbers,amsmath,amssymb]{revtex4}

\usepackage{epsfig}

\input minisym

\begin{document}

\preprint{\babar-PUB-05/05}
\preprint{SLAC-PUB-11896}

\title{Measurement of the Spin of the $\Omega^-$ Hyperon at \babar}

%
\author{B.~Aubert}
\author{R.~Barate}
\author{M.~Bona}
\author{D.~Boutigny}
\author{F.~Couderc}
\author{Y.~Karyotakis}
\author{J.~P.~Lees}
\author{V.~Poireau}
\author{V.~Tisserand}
\author{A.~Zghiche}
\affiliation{Laboratoire de Physique des Particules, F-74941 Annecy-le-Vieux, France }
% ...
\collaboration{The \babar\ Collaboration}
\noaffiliation

\date{\today}

\begin{abstract}
A measurement of the spin of the $\Omega^-$ hyperon produced through the exclusive process $\Xi_c^0 \rightarrow \Omega^- K^+$ 
is presented using a total integrated luminosity of 116 $\rm{fb}^{-1}$ recorded with the
\babar\ detector at the $e^+ e^-$ asymmetric-energy $B$-Factory at SLAC.  
Under the assumption that the $\Xi_c^0$ has spin 1/2, the angular distribution of the 
$\Lambda$ from $\Omega^- \rightarrow \Lambda K^-$ decay is inconsistent with all
half-integer $\Omega^-$ spin values other than 3/2.
Lower statistics data for the process 
$\Omega_c^0 \rightarrow \Omega^- \pi^+$ from a 230 $\rm{fb}^{-1}$ sample 
are also found to be consistent with $\Omega^-$ spin 3/2.  If the 
$\Xi_c^0$ spin were 3/2, an $\Omega^-$ spin of 5/2 cannot be excluded.

\end{abstract}

\maketitle

The $SU(3)$ classification scheme predicted~\cite{ref:GellMann} the existence of the $\Omega^{-}$ hyperon, an
isosinglet with hypercharge $Y=-2$ and strangeness $S=-3$, as a member
of the $J^{P} = 3/2^{+}$ ground state baryon decuplet. 
Such a particle was observed subsequently with the predicted mass in a bubble chamber experiment~\cite{ref:Barnes}.  
In previous attempts to confirm the spin of the $\Omega^{-}$~\cite{ref:Deutschmann, ref:Baubillier, ref:cern}, 
$K^-$ $p$ interactions in a liquid hydrogen bubble chamber were studied.   
In each case only a small $\Omega^{-}$ data sample was obtained, and the 
$\Omega^{-}$ production mechanism was not well understood.  As a result, 
these experiments succeeded only in establishing that the $\Omega^-$ spin 
is greater than 1/2.

In this letter, measurements of the $\Omega^-$ spin are obtained using 
$\Omega^-$ samples~\cite{ref:ChConj} from the decay of $\Xi_{c}^{0}$ and $\Omega_{c}^{0}$ charm baryons 
inclusively produced in $e^+e^-$ collisions at center-of-mass energies 10.58 and 10.54 GeV.  
The primary $\Omega^-$ sample is obtained from the decay sequence $\Xi_{c}^{0} \rightarrow \Omega^{-} K^+$, with 
$\Omega^{-} \rightarrow \Lambda K^-$, while a much smaller sample  
resulting from $\Omega_{c}^{0} \rightarrow \Omega^{-} \pi^+$, with
$\Omega^{-} \rightarrow \Lambda K^-$ is used for corroboration.
It is assumed that each charm baryon type has spin 1/2 and, as a result 
of its inclusive production, that it is described by a diagonal spin projection 
density matrix.  The analysis does not require that the diagonal matrix elements be equal.  

The helicity 
formalism~\cite{ref:form, ref:form2} is applied in order to examine the implications of various $\Omega^-$ spin hypotheses  
for the angular distribution of the $\Lambda$ from $\Omega^-$ decay. 
\begin{figure}
  \centering\small
    \includegraphics[width=0.4\textwidth]{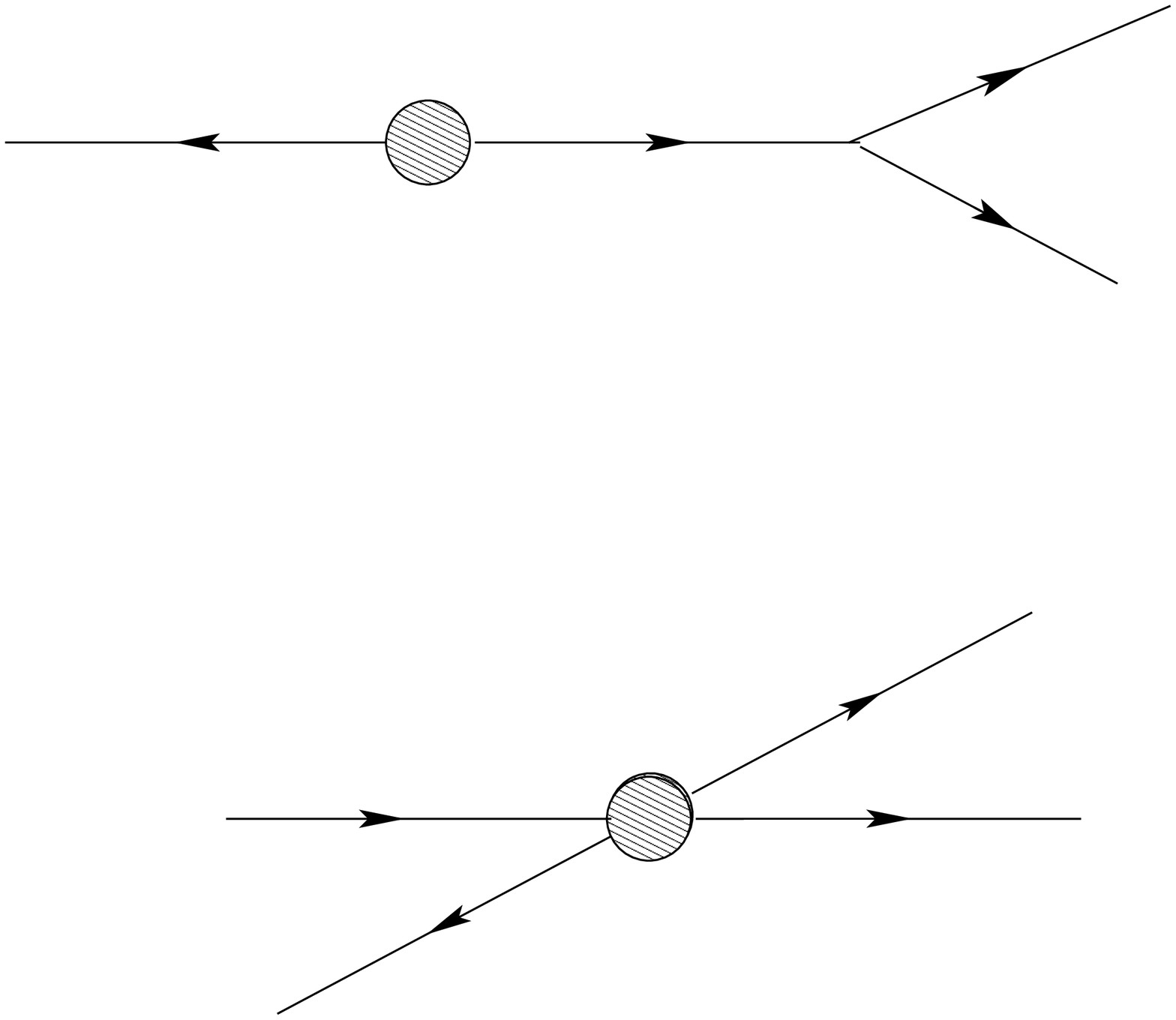} 
    \begin{picture}(0.,0.)
    \put(-205,190){$\vec{K}^+_1$}
    \put(-150,185){$\vec{\Xi}^0_c=\vec{0}$}
    \put(-81,190){$\vec{\Omega}^-_1$}
    \put(0,228){$\vec{\Lambda}_1$}
    \put(-10,172){$\vec{K}^-_1$}
    \put(-210,160){a) All decay products in the $\Xi^0_c$ rest-frame.}
    \put(-185,85){$\vec{\Omega}^-_1$}
    \put(-110,66){$\vec{\Omega}^-_2=\vec{0}$}
    \put(-80,90){)}
    \put(-68,92){\bf{$\theta_h$}}
    \put(-22,122){$\vec{\Lambda}_2$}
    \put(-178,45){$\vec{K}^-_2$}
    \put(-210,20){b) All decay products in the $\Omega^-$ rest-frame;}
    \put(-200,5){   in this frame, $\vec{\Omega}^-_1 \rightarrow \vec{\Omega}^-_2=\vec{0}$, $\vec{\Lambda}_1 \rightarrow \vec{\Lambda}_2$, $\vec{K}^-_1 \rightarrow \vec{K}^-_2$.}
    \end{picture}
  \caption{Schematic definition of the helicity angle $\theta_{h}$ in the decay chain $\Xi_c^0 \rightarrow \Omega^- K^+$, $\Omega^- \rightarrow \Lambda K^-$; as shown in b) $\theta_{h}$ is the angle between the $\Lambda$ direction in the $\Omega^{-}$ rest-frame and the $\Omega^{-}$ direction in the $\Xi_c^0$ rest-frame (the quantization axis).}
  \label{fig:Helicity}
\end{figure}
By choosing the quantization axis along 
the direction of the $\Omega^-$ in the charm baryon rest-frame, the $\Omega^-$ inherits the spin 
projection of the charm baryon, since any orbital angular momentum in the charm baryon decay has no projection in this direction.  
It follows that, regardless of the spin $J$ of the $\Omega^-$, the density matrix describing the $\Omega^-$ sample is 
diagonal, with non-zero values only for the $\pm 1/2$ spin projection elements,  
i.e. the helicity $\lambda_i$ of the $\Omega^-$ can take only the values  $\pm 1/2$.  Since the final state $\Lambda$ and $K^-$
have spin values 1/2 and 0, respectively, the net final state helicity $\lambda_f$ also can take only the values $\pm 1/2$.  
The helicity angle $\theta_h$ is then defined as the angle between the direction of the $\Lambda$ in the rest-frame of the $\Omega^-$ 
and the quantization axis (Fig.~\ref{fig:Helicity}).

The probability for the $\Lambda$ to be produced with Euler angles $(\phi, \theta_h, 0)$ with respect to the quantization axis
is given by the square of the amplitude $\psi$,  characterizing the decay of an $\Omega^-$ with
total angular momentum $J$ and helicity $\lambda_{i}$
to a 2-body system with net helicity $\lambda_{f}$,
\begin{eqnarray}
\psi = A^J_{\lambda_f} D^{J *}_{\lambda_{i} \lambda_{f}}(\phi, \theta_{h}, 0), 
\end{eqnarray}
where the transition matrix element $A^J_{\lambda_f}$ 
represents the coupling of the $\Omega^-$ to the final state, 
and $D^{J}_{\lambda_{i} \lambda_{f}}$ is an element of the Wigner rotation matrix~\cite{ref:form3}; 
$A^J_{\lambda_f}$ does not depend on $\lambda_i$ because of 
rotational invariance (Wigner-Eckart theorem~\cite{ref:Eckart}).
The angular distribution of the $\Lambda$ is then given by the total intensity,
\begin{eqnarray}
I\propto \sum_{\lambda_{i}, \lambda_{f}} \rho_{i}\left |A^J_{\lambda_f}D^{J *}_{\lambda_{i} \lambda_{f}}(\phi, \theta_h, 0)\right |^2, 
\end{eqnarray}
\noindent where the $\rho_i$ ($i= \pm 1/2$) are the diagonal density matrix elements 
inherited from the charm baryon, and the sum is over all initial and final helicity states.

Using this expression, the $\Lambda$ angular distribution integrated 
over $\phi$ is obtained for spin hypotheses $J_{\Omega}=1/2$, $3/2,$ and $5/2$, respectively as follows:
\begin{eqnarray}
{dN}/{d\rm{cos} \theta_{\it h}}&\propto& 1+\beta\, \rm {cos}\theta_{\it h}\\
{dN}/{d\rm{cos} \theta_{\it h}}&\propto& 1 + 3\,{\rm cos}^2\theta_{\it h}+\beta\, \rm {cos}\theta_{\it h}(5- 9\,\rm cos^2\theta_{\it h}) \;\;\;\;\;\;\\
\nonumber {dN}/{d\rm{cos} \theta_{\it h}}&\propto&  1-2\,{\rm cos}^2\theta_{h}+5\,{\rm cos}^4\theta_{\it h} \\
&&+\beta\, \rm {cos}\theta_{h}(5-26\,{\rm cos}^2\theta_{\it h}+25\,{\rm cos}^4\theta_{\it h}), 
\end{eqnarray}
where the coefficient of the asymmetric term 
$$\beta=\left[\frac{\rho_{1/2}-\rho_{-1/2}}{\rho_{1/2}+\rho_{-1/2}}\right] \left[ {\frac{\left |A^J_{1/2}\right |^2-\left |A^J_{-1/2}\right |^2}{\left |A^J_{1/2}\right |^2+\left |A^J_{-1/2}\right |^2 }}\right ] $$ 
may be non-zero as a consequence of parity violation in charm baryon and $\Omega^-$ weak decay.   
Eqs.~(3) and (4) are the distributions considered in connection with the discovery of the $\Delta(1232)$ 
resonance~\cite{ref:fermi}, generalized to account for parity violation.

The data samples used for this analysis were collected with the \babar\ detector
at the \pep2\ asymmetric energy $e^+e^-$ collider and correspond to a total integrated luminosity of
116 fb$^{-1}$ and 230 fb$^{-1}$ for the $\Xi_{c}^{0} \rightarrow \Omega^- \: K^+$
and $\Omega_{c}^{0} \rightarrow \Omega^- \: \pi^+$ samples, respectively.
The detector is described
in detail elsewhere~\cite{ref:babar}.
The selection of $\Xi_c^0$ and $\Omega_{c}^{0}$ candidates requires
the intermediate reconstruction of events consistent with 
$\Omega^-   \rightarrow \Lambda \: K^-$  
and $\Lambda \rightarrow p \: \pi^-$. Particle identification selectors for the proton and the kaons, 
based on specific energy loss (${\rm d}E/{\rm d}x$) and Cherenkov angle measurements, have been used~\cite{ref:babar}. 
Each intermediate state candidate 
is required to have its
invariant mass within a $\pm 3\sigma$ mass window centered on the fitted peak position of the relevant distribution,
where $\sigma$ is the mass resolution obtained from the fit.  
In all cases, the fitted peak mass is consistent with the expected value~\cite{ref:pdg}.
The intermediate state invariant mass is then constrained to its nominal value~\cite{ref:pdg}.

Since the hyperons are long-lived,
the signal-to-background ratio is improved by imposing vertex displacement criteria.
The distance between the  $\Omega^- K^+$ or $\Omega^- \pi^+$ vertex and the  $\Omega^-$  
decay vertex, when projected onto the plane perpendicular to the collision axis, must 
exceed 1.5~mm in the $\Omega^-$ direction.  
The distance between the $\Omega^-$ and $\Lambda$ decay
vertices is required to exceed 1.5~mm in the direction of the
$\Lambda$ momentum vector.
In order to further enhance signal-to-background ratio, a selection criterion is imposed on the 
center-of-mass momentum $p^*$ of the charm baryon: 
$p^* > 1.8$~GeV$/c$ for $\Xi_c^0$ and $p^* > 2.5$~GeV$/c$ for $\Omega_c^0$ candidates.
In addition, a minimum laboratory momentum requirement of 200 MeV$/c$ is imposed on the $\pi^+$ daughter of the $\Omega_{c}^{0}$ in order to 
reduce combinatorial background level due to soft pions.
The invariant mass spectra of $\Xi_{c}^{0}$ and $\Omega_{c}^{0}$ candidates in data are shown 
before efficiency correction in Figs.~2(a) and~2(b), respectively.
The signal yields (770$\pm$33  $\Xi_{c}^{0}$ and 159$\pm$17  $\Omega_{c}^{0}$ candidates) are obtained 
from fits with a double Gaussian ($\Xi_{c}^{0}$) or single Gaussian ($\Omega_{c}^{0}$) 
signal function and a linear background function.  The corresponding selection efficiencies obtained from 
Monte Carlo simulations are $14.7\%$ and $15.8\%$, respectively.

\begin{figure}
  \centering\small
    \includegraphics[width=0.35\textwidth]{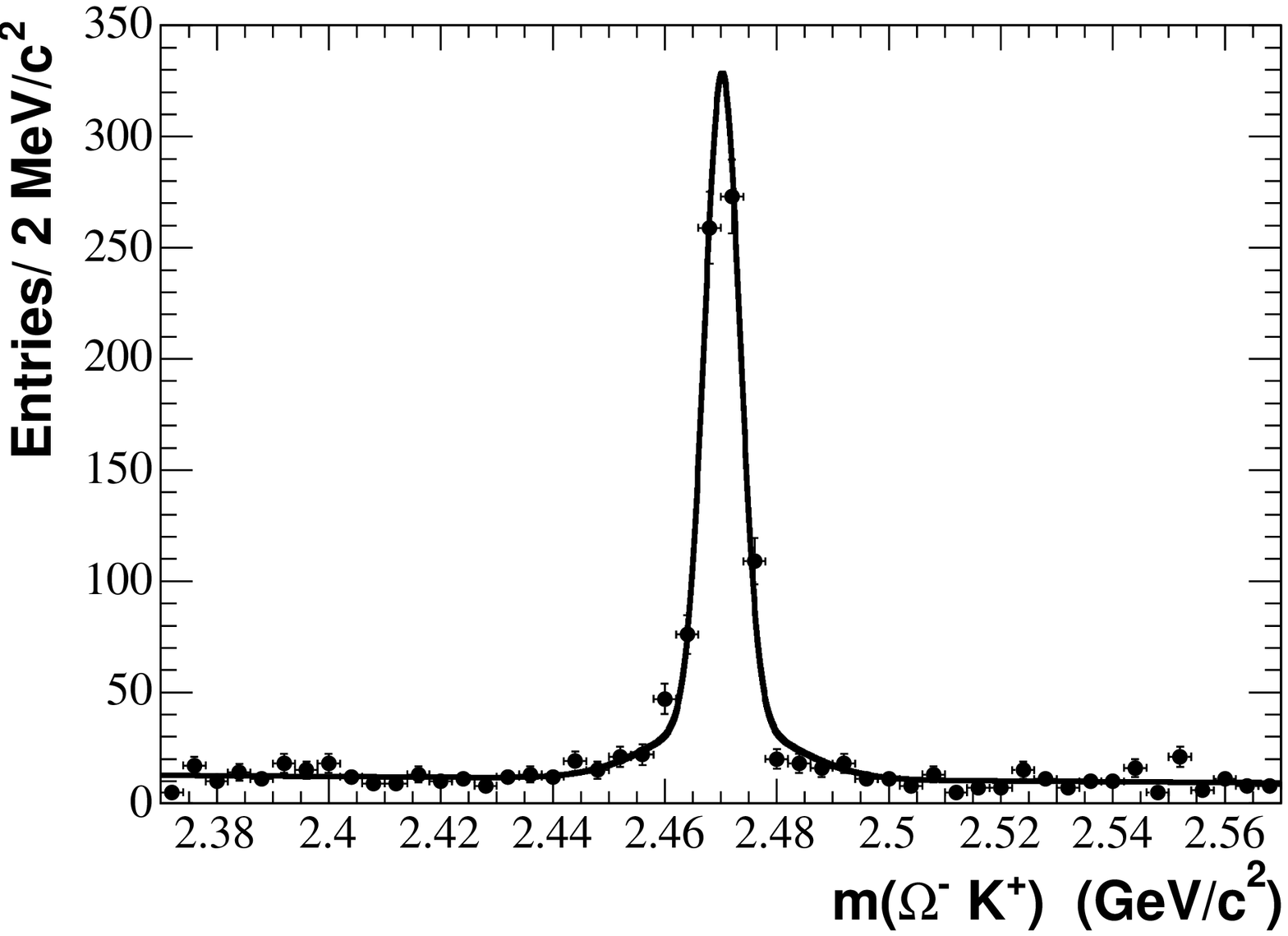} \\
    \includegraphics[width=0.35\textwidth]{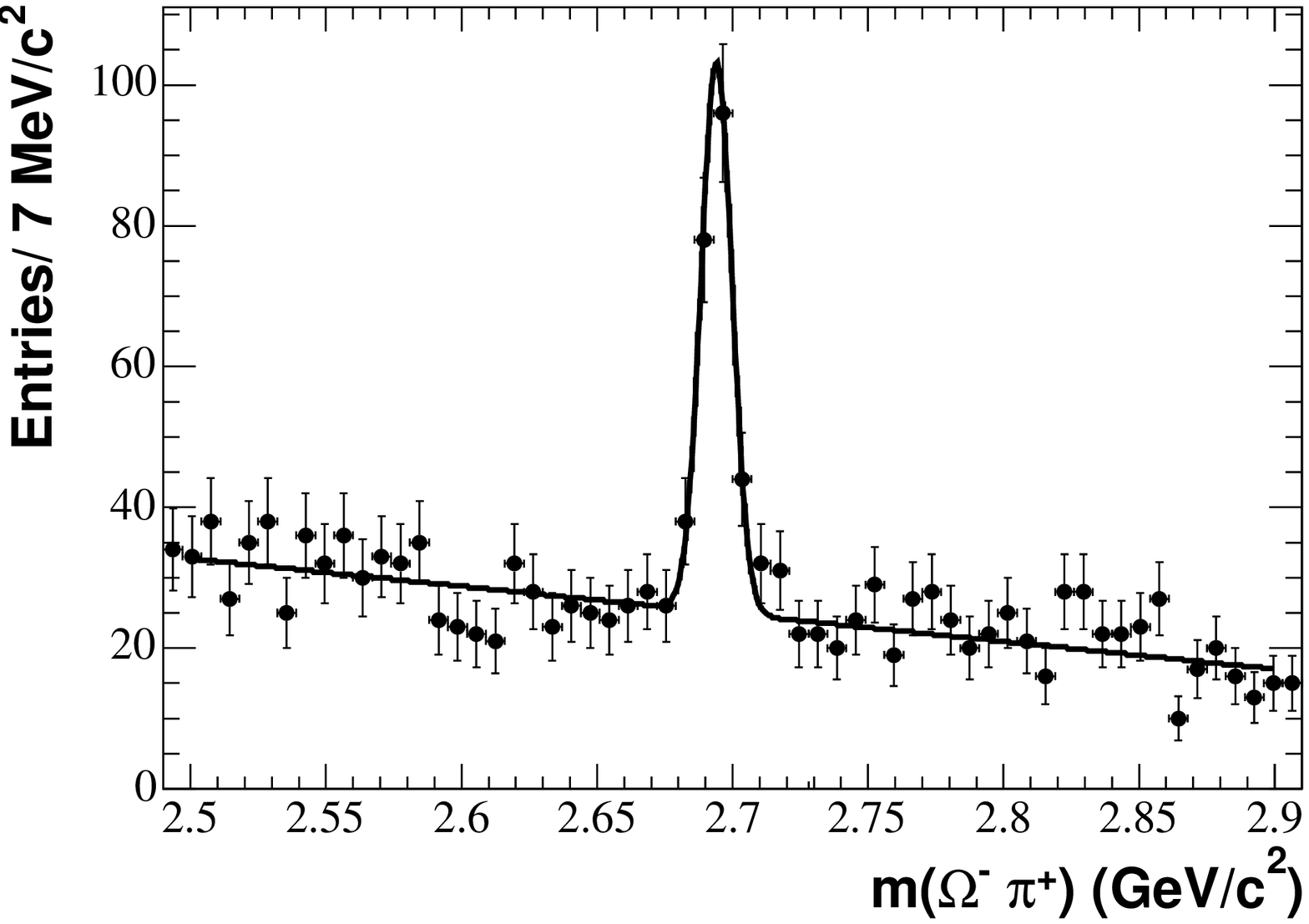}
    \begin{picture}(0.,0.)
    \put(-155,240){\bf{(a)}}
    \put(-155,110){\bf{(b)}}
    \end{picture}
  \caption[The uncorrected $\Omega^- K^+$ (a) and $\Omega^- \pi^+$ (b) invariant mass spectra in data. 
The curves result from the fits described in the text.]
          {The uncorrected $\Omega^- K^+$ (a) and $\Omega^- \pi^+$ (b) invariant mass spectra in data. 
The curves result from the fits described in the text.}
  \label{fig:Xicmass}
\end{figure}

\begin{figure}
  \centering\small
    \includegraphics[width=0.35\textwidth]{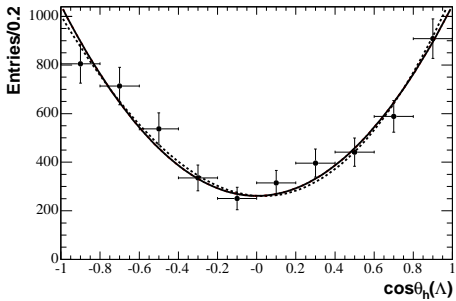} 
    \begin{picture}(0.,0.)
    \end{picture}
\caption[The efficiency-corrected cos$\theta_h(\Lambda)$ distribution for $\Xi_c^0 \rightarrow \Omega^- K^+$ data.
The dashed curve shows the $J_\Omega=3/2$ fit using Eq.~(4), 
in which $\beta$ allows for possible asymmetry. 
The solid curve represents the corresponding fit with $\beta=0$.]
{The efficiency-corrected cos$\theta_h(\Lambda)$ distribution for $\Xi_c^0 \rightarrow \Omega^- K^+$ data.
The dashed curve shows the $J_\Omega=3/2$ fit using Eq.~(4), 
in which $\beta$ allows for possible asymmetry. 
The solid curve represents the corresponding fit with $\beta=0$.}
\label{fig:CosAsymm}
\end{figure}

For the $\Omega^-$ sample resulting from $\Xi_c^0$ decay, 
the uncorrected cos$\theta_h(\Lambda)$ distribution is obtained 
by means of an unbinned maximum likelihood fit to the $\Omega^- K^+$ invariant mass spectrum
corresponding to each of ten equal intervals of cos$\theta_h(\Lambda)$ in the range $-1$ to $1$.
In each interval the $\Xi_c^0$ signal function shape is fixed to that obtained from the fit 
shown in Fig.~\ref{fig:Xicmass}~(a).  
The $\Xi_c^0$ reconstruction efficiency in each interval of cos$\theta_h(\Lambda)$ is 
obtained from Monte Carlo simulation, and the resulting  efficiency-corrected distribution is shown in Fig.~\ref{fig:CosAsymm}.
The measured efficiency varies linearly from 14.0\% at cos$\theta_h(\Lambda)=-1$ 
to 15.3\% at cos$\theta_h(\Lambda)=+1$, and so the shape of the angular distribution is changed 
only slightly by the correction procedure.  
The dashed curve corresponds to a fit of the $J_\Omega=3/2$ parametrization of Eq.~(4) and yields $\beta=0.04\pm 0.06$. 
The forward-backward asymmetry $A=(F-B)/(F+B)$ of the efficiency-corrected cos$\theta_h(\Lambda)$ distribution of Fig.~3, 
where $B$ ($F$) represents the number of signal events satisfying 
cos$\theta_h(\Lambda)\leq 0$ ($\geq 0$), is $+0.001 \pm 0.019$.  
This and the fitted value of $\beta$ indicate that the data show no significant asymmetry, and so we set $\beta=0$ 
in subsequent fits.
The solid curve represents the fit to the data with $\beta=0$; 
the fit information relevant to Eq.~(4) is indicated in Table 1.  

\begin{figure}
  \centering\small
    \includegraphics[width=0.35\textwidth]{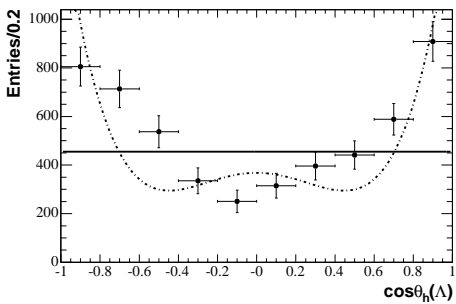}
    \begin{picture}(0.,0.)
    \end{picture}
\caption[The efficiency-corrected cos$\theta_h(\Lambda)$ distribution for $\Xi_c^0 \rightarrow \Omega^- K^+$ data. 
The solid line represents the expected distribution for $J_{\Omega}=1/2$, while 
the dashed curve corresponds to $J_{\Omega}=5/2$. In each case, $\beta=0$.]
{The efficiency-corrected cos$\theta_h(\Lambda)$ distribution for $\Xi_c^0 \rightarrow \Omega^- K^+$ data. 
The solid line represents the expected distribution for $J_{\Omega}=1/2$, while 
the dashed curve corresponds to $J_{\Omega}=5/2$. In each case, $\beta=0$.}
  \label{fig:CosXic}
\end{figure}

The efficiency-corrected cos$\theta_h(\Lambda)$ distribution with fits corresponding to 
Eqs.~(3) and (5) with $\beta=0$ is shown in Fig.~\ref{fig:CosXic}. 
The solid line represents the expected distribution for $J_{\Omega}=1/2$, while 
the dashed curve corresponds to $J_{\Omega}=5/2$.
The corresponding values of fit confidence level (C.L.) are extremely small (Table 1).
\begin{table}
\begin{center}
\begin{tabular}{c c r c l c l }
\hline \hline 
  $J_\Omega$   & & Fit $\chi^2/$NDF & & Fit C.L. & & Comment\\ \hline 
   1/2    & &  100.4/9  & &    $1\times 10^{-17}$ & & Fig.~4, solid line  \\  
   3/2    & &  6.5/9    & &   0.69 ($\beta = 0$) & & Fig.~3, solid curve\\
   3/2    & &  6.1/8    & &   0.64 ($\beta\neq 0$) & & Fig.~3, dashed curve\\ 
   5/2    & &  47.6/9  & &   $3\times 10^{-7}$   & & Fig.~4, dashed curve\\  \hline \hline
\end{tabular}
\end{center}
\caption[The ${\rm cos}\theta_{h}(\Lambda)$ angular distribution fit C.L. values corresponding to $\Omega^-$ spin hypotheses 1/2, 3/2 and 5/2 
for  $\Xi_{c}^{0}\rightarrow \Omega^- K^+$ data assuming $J_{\Xi_c}=1/2$.]
{The ${\rm cos}\theta_{h}(\Lambda)$ angular distribution fit C.L. values corresponding to $\Omega^-$ spin hypotheses 1/2, 3/2 and 5/2 
for  $\Xi_{c}^{0}\rightarrow \Omega^- K^+$ data assuming $J_{\Xi_c}=1/2$.}
\label{tab:xicfits}
\end{table}
For $J_\Omega\ge 7/2$, the predicted angular distribution increases even more steeply for $|\rm cos\theta_{h}|\sim 1$ 
than for $J_\Omega = 5/2$ and exhibits ($2J_\Omega -2$) turning points.  The relevant fit C.L. values are even 
smaller than that for $J_\Omega = 5/2$, and so $J_\Omega\ge 7/2$ can be excluded at C.L. greater than 99\%.

\begin{figure} 
  \centering\small
    \includegraphics[width=0.35\textwidth]{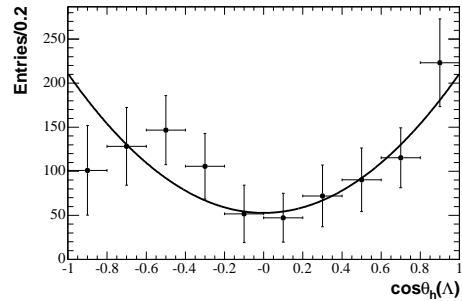} 
    \begin{picture}(0.,0.)
    \end{picture}
  \caption[The efficiency-corrected cos$\theta_{h}(\Lambda)$
distribution in data for $\Omega_{c}^{0}\rightarrow \Omega^- \pi^+$ events.  
The curve corresponds to $J_{\Omega_c}=1/2$ and $J_{\Omega^-}=3/2$ with $\beta=0$.]
          {The efficiency-corrected cos$\theta_{h}(\Lambda)$
distribution in data for $\Omega_{c}^{0}\rightarrow \Omega^- \pi^+$ events.  The curve corresponds to $J_{\Omega_c}=1/2$ and $J_{\Omega^-}=3/2$ with $\beta=0$.}
  \label{fig:Result_Lambda_hel_Eff_Corr_omegac}
\end{figure}

These fit results were checked using the sample of $\Omega^{-}$ hyperons obtained from $\Omega_c$ baryon decays.
The $\Omega_c$ baryon is presumed to belong to the $\bf 6$ representation of an $SU(3)$ $J^{P}=1/2^{+}$ multiplet~\cite{ref:pdg}, so 
that the $\Omega^-$ decay angular distribution 
should again be proportional to $(1 + 3\,{\rm cos}^2\theta_{h})$.  
After efficiency-correction, the 
angular distribution shown in Fig.~\ref{fig:Result_Lambda_hel_Eff_Corr_omegac} is found to be consistent with 
$J_{\Omega^-}=3/2$ with $\beta$ again set to zero.
The fit to the corrected distribution has $\chi^2$/NDF = 6.5/9 and C.L. 0.69, and so 
is in very good agreement with the results obtained from $\Xi_c^0$ decay.
The fit for $\beta$ yields $\beta=0.4\pm 0.2$ and the value of the forward-backward asymmetry is $+0.013\pm 0.058$.
 
The implications for the spin of the $\Omega^-$ if the spin of the 
$\Xi_c^0$ is assumed to be 3/2 are now considered.  
For $J_{\Omega}=1/2$, the predicted decay angular distribution is again 
given by Eq.~(3), and so this possibility can be ruled out.

If asymmetric contributions are ignored, the $\Omega^-$ angular distribution for spin 
values 3/2 and 5/2 are determined by the values of the quantities $x=\rho_{3/2}+\rho_{-3/2}$ 
and $(1-x) = \rho_{1/2}+\rho_{-1/2}$.  
For $J_{\Omega}=3/2$, $x=0$ would yield a distribution given by Eq.~(4) with $\beta =0$, in excellent 
agreement with the data. 
 However, for inclusive $\Xi_c^0$
production with the $\Omega^-$ direction in the $\Xi_c^0$ rest-frame as
quantization axis, it would seem more reasonable to expect the spin
projection states to be populated equally. This would yield $x = 0.5$,
and would result in an isotropic $\Omega^-$ decay distribution, in clear
disagreement with the observed behavior. 

A consequence of such a
$\Xi_c^0$ density matrix configuration would be that there should be
no preferred direction in the decay to $\Omega^- K^+$ in the $\Xi_c^0$
rest-frame.  This hypothesis has been tested 
in the present analysis by measuring the $\Xi_c^0$
polarization with respect to its production-plane normal; there is no
evidence for such polarization.  In addition, the
spherical harmonic ($Y_L^M$) moments of the $\Xi_c^0$ decay angular
distribution for $L\leq$ 6 and $M \leq 6$ have been compared 
to those obtained from  
simulation in which the $\Xi_c^0$ decay is isotropic; no significant
difference was found. It is therefore reasonable to infer that
the combination $J_{\Xi_c} = 3/2$ and $J_{\Omega}=3/2$ is disfavored.

For $J_{\Omega}=5/2$ the situation is quite different. The decay
angular distribution is then
\begin{eqnarray}
\nonumber {dN}/{d\rm{cos} \theta_{\it h}}& \propto& 10{\rm cos}^4\theta_{\it h} - 4{\rm cos}^2\theta_{\it h} + 2 \\ 
                & & - {\it x} (25{\rm cos}^4\theta_{\it h} - 18{\rm cos}^2\theta_{\it h} + 1). 
\end{eqnarray} 
In this case, $x = 0.5$ gives 
\begin{eqnarray}
{dN}/{d\rm{cos} \theta_{\it h}} \propto -5{\rm cos}^4\theta_{\it h} + 10{\rm cos}^2\theta_{\it h} + 3,
\end{eqnarray} 
which has a minimum at cos$\theta_h = 0$, maxima at cos$\theta_h =\pm 1$,
and fits the observed angular distribution with C.L. 0.44. If
 $x$ is allowed to vary, the best fit to the data has $x = 0.4$, which
corresponds to 
\begin{eqnarray}
{dN}/{d\rm{cos} \theta_{\it h}} \propto 1 + 2{\rm cos}^2\theta_{\it h}\, ; 
\end{eqnarray}
 the quartic term is thus cancelled, and fit C.L. 0.53 is
obtained.
 
   It follows from this discussion that for $J_{\Xi_c} = 3/2$, the
hypothesis $J_{\Omega}=1/2$ is ruled out, and $J_{\Omega}=3/2$  may 
reasonably be considered disfavored; however, $J_{\Omega}=5/2$
is entirely acceptable. For this reason, it has been emphasized 
that the determination that the $\Omega^-$
has spin 3/2 is entirely contingent upon the assumption that the spin
of the $\Xi_c^0$ (and of the $\Omega_c^0$) is 1/2.

In conclusion, the angular distributions of the decay products of the $\Omega^{-}$ baryon resulting from
 $\Xi_c^0$ and $\Omega_c^0$ decays are well-described 
by a function $\propto(1 + 3{\rm cos}^2\theta_{h})$.  These observations are consistent with spin assignments
1/2 for the $\Xi_c^0$ and the $\Omega_c^0$, and 3/2 for the $\Omega^-$.
Values of 1/2 and  greater than 3/2 for the spin of the $\Omega^{-}$ 
yield C.L. values significantly less than 1\% when spin 1/2 is assumed for the parent charm baryon.

\begin{acknowledgments}

We are grateful for the excellent luminosity and machine conditions
provided by our \pep2\ colleagues, 
and for the substantial dedicated effort from
the computing organizations that support \babar.
The collaborating institutions wish to thank 
SLAC for its support and kind hospitality. 
This work is supported by
DOE
and NSF (USA),
NSERC (Canada),
IHEP (China),
CEA and
CNRS-IN2P3
(France),
BMBF and DFG
(Germany),
INFN (Italy),
FOM (The Netherlands),
NFR (Norway),
MIST (Russia), and
PPARC (United Kingdom). 
Individuals have received support from CONACyT (Mexico), A.~P.~Sloan Foundation, 
Research Corporation,
and Alexander von Humboldt Foundation.

\end{acknowledgments}


\begin{thebibliography} {19}
\bibitem{ref:ChConj}
The use of charge conjugate states is implied throughout.
\end{thebibliography}
\end{document}